\documentclass[a4paper]{article}

\def\squareforendex{\hbox{$\lozenge$}}
\def\endex{\ifmmode\squareforendex\else{\unskip\nobreak\hfil
\penalty50\hskip1em\null\nobreak\hfil\squareforendex
\parfillskip=0pt\finalhyphendemerits=0\endgraf}\fi}

\usepackage{amsmath,amssymb,amsthm}
\usepackage{dsfont}
\usepackage{hyperref}

\usepackage{triotex}
\usepackage{times}

\newcommand{\vari}[1]{\mathbf{var}\!\left({#1}\right)}

\newcommand{\nfy}{\varkappa\langle y \rangle}
\newcommand{\nfzpn}{\varkappa\langle z,\mathsf{Pn}\rangle}
\newcommand{\nfzseq}{\varkappa\langle z,\mathsf{U} \rangle}
\newcommand{\nfzpi}{\varkappa\langle x,\pi \rangle}

\newcommand{\ns}{\neg s}

\newcommand{\kbb}{\mathbb{K}}
\newcommand{\dbb}{\mathbb{D}}
\newcommand{\vbb}{\mathbb{V}}
\newcommand{\ibb}{\mathbb{I}}
\newcommand{\jbb}{\mathbb{J}}

\newcommand{\etap}[1]{\ifthenelse{\equal{#1}{1}} {\eta^{\ref{eq:normalform}}} {\ifthenelse{\equal{#1}{2}} {\eta^{\ref{eq:normalform-G}}} {\eta^{\ref{eq:normalform-X}}}}}

\DeclareRobustCommand{\form}[1]%
{\ensuremath{%
	 \ifthenelse{ \not \equal{#1}{} } {\mathtt{L}\!\left(#1\right)} {\mathtt{L}} }
}%

\DeclareRobustCommand{\lang}[1]%
{\ensuremath{%
	 \ifthenelse{ \not \equal{#1}{} } {\mathbf{L}\!\left(#1\right)} {\mathbf{L}} }
}%

\DeclareRobustCommand{\formk}[2]{\ifthenelse{ \not \equal{#1}{} \AND \not \equal{#2}{} }{\form{\mathsf{U}^{#1}, \mathsf{X}^{#2}}}{\ifthenelse{ \not \equal{#1}{} }{\form{\mathsf{U}^{#1}, \mathsf{X}}}{\ifthenelse{ \not \equal{#2}{} }{\form{\mathsf{U}, \mathsf{X}^{#2}}}{\form{\mathsf{U}, \mathsf{X}}}}}}

\DeclareRobustCommand{\langk}[2]{\ifthenelse{ \not \equal{#1}{} \AND \not \equal{#2}{} }{\lang{\mathsf{U}^{#1}, \mathsf{X}^{#2}}}{\ifthenelse{ \not \equal{#1}{} }{\lang{\mathsf{U}^{#1}, \mathsf{X}}}{\ifthenelse{ \not \equal{#2}{} }{\lang{\mathsf{U}, \mathsf{X}^{#2}}}{\lang{\mathsf{U}, \mathsf{X}}}}}}

\newcommand{\sizeU}[1]{|#1|_{\mathsf{U}}}

\newcommand{\sizep}[1]{|#1|_{\mathsf{p}}}
\newcommand{\sizeX}[1]{|#1|_{\mathsf{X}}}
\newcommand{\dist}[1]{d(#1)}
\newcommand{\sub}[1]{s(#1)}

\newcommand{\Pn}[3]{\MTLoperator{\mathsf{Pn}^{#1}_{#2}}{}{}{#3}}

\newcommand{\exPn}[4]{\MTLoperator{\mathsf{exqPn}^{{#1};\langle #2 \rangle}_{#3}}{}{}{#4}}

\newcommand{\frf}[1]{(\ref{#1})}
\newcommand{\fsrf}[2]{(\ref{#1}--\ref{#2})}

\newcommand{\statech}[1]{\curlywedge(#1)}
\newcommand{\statechm}[1]{\curlyvee(#1)}
\newcommand{\relaxX}[2]{\mathcal{X}^{#1}(#2)}
\newcommand{\relaxOne}[1]{\mathcal{U}(#1)}
\newcommand{\relaxTwo}[1]{\mathcal{R}(#1)}
\newcommand{\Xcal}{\mathcal{X}}

\newcommand{\ltl}{\ensuremath{\mathtt{LTL}}}

\newcommand{\expnltl}{\ensuremath{\mathtt{LTL}(\mathsf{U}, \mathsf{exqPn})}}

\newcommand{\qltl}{\form{\mathsf{U}}}
\newcommand{\propos}[1]{\mathtt{P}(#1)}

\newcommand{\logictrue}{\top}
\newcommand{\logicfalse}{\bot}

\newcommand{\Scal}{\mathcal{S}}
\newcommand{\Pcal}{\mathcal{P}}
\newcommand{\Qcal}{\mathcal{Q}}
\newcommand{\Rcal}{\mathcal{M}}
\newcommand{\Hcal}{\mathcal{H}}

\newcommand{\Orm}{\mathrm{O}}

\newcommand{\allmodels}[1]{\ldoubsq{#1}\rdoubsq}
\newcommand{\allwords}[1]{\mathcal{W}[{#1}]}

\newcommand{\range}[2]{{#1}\!\!:\!\!{#2}}

\theoremstyle{plain}
\newtheorem{theorem}{Theorem}
\newtheorem{proposition}[theorem]{Proposition}
\newtheorem{lemma}[theorem]{Lemma}

\theoremstyle{definition}

\newtheorem{example}[theorem]{Example}

\title{\textsc{On Relaxing Metric Information \\ in Linear Temporal Logic}}
\author{Carlo A. Furia\footnote{Chair of Software Engineering, ETH Z\"urich, Switzerland. E-mail: \texttt{caf@inf.ethz.ch}.} $\quad$and$\ \,$ Paola Spoletini\footnote{Universit\`a degli Studi dell'Insubria, Italy. E-mail: \texttt{paola.spoletini@uninsubria.it}.}}

\begin{document}

\maketitle

\begin{abstract}
Metric LTL formulas rely on the \emph{next} operator to encode time distances, whereas qualitative LTL formulas use only the \emph{until} operator.
This paper shows how to transform any metric LTL formula $M$ into a qualitative formula $Q$, such that $Q$ is satisfiable if and only if $M$ is satisfiable over words with variability bounded with respect to the largest distances used in $M$ (i.e., occurrences of \emph{next}), but the size of $Q$ is independent of such distances.
Besides the theoretical interest, this result can help simplify the verification of systems with time-granularity heterogeneity, where large distances are required to express the coarse-grain dynamics in terms of fine-grain time units.
\end{abstract}

\section{Introduction and motivation}
Linear temporal logic (\ltl{}) supports a simple model of \emph{metric time} through the \emph{next} operator $\Xltl$.
Under the assumption of a one-to-one correspondence between consecutive states and discrete instants of time, nested occurrences of $\Xltl$ ``count'' instants to express time distances.
\ltl{} formulas without $\Xltl$ --- using only the \emph{until} operator --- are instead purely \emph{qualitative}: they constrain the ordering of events, not their absolute distance.
Therefore, qualitative \ltl{} formulas express models that are insensitive to additions or removals of \emph{stuttering steps}: consecutive repetitions of the same state.
The fundamental properties of \ltl{} with respect to its qualitative subset are well known from classic work: quantitative (metric) \ltl{} is strictly more expressive \cite{Lam83,PW97,Ete00,KS05}, but reasoning has the same worst-case complexity \cite{SC85,DS02}.

The present paper investigates when the metric information, encoded by nested occurrences of $\Xltl$, is redundant and can be relaxed.
The relaxation transforms a quantitative formula into an equi-satisfiable qualitative one that is independent of the number of $\Xltl$ in the original formula; reasoning on the transformed formula is thus simpler by a factor proportional to the amount of metric information stripped.

%The present paper studies instead the \emph{equi-satisfiability} of  and qualitative \ltl{} formulas and determines under what conditions the satisfiability problem for \ltl{} is reducible from metric to qualitative.
The motivation behind this study refers to an informal notion of \emph{redundancy}, which stuttering steps seem to encode.
Consider a metric \ltl{} formula $\phi$ describing models characterized by many stuttering steps distributed over large time distances; for example, the formalization of an event for elections that occur every four years in November, in a variable day of the month, with the day as time unit.
Formula $\phi$ is large because it encodes large time distances in unary form with many occurrences of the $\Xltl$ operator; for example, a four-year distance requires at least $1460$ ``next'', one for each day.
However, the information carried by $\phi$ is prominently redundant as every stuttering step is a duplication that only pads uneventful time instants.
Is it possible, under a rigorous assumption of ``sparse events'', to simplify $\phi$ into an equi-satisfiable formula $\phi'$ which does not encode explicitly the redundant information?

The notion of \emph{bounded variability}, adapted from dense-time models, provides a suitable formalization of the intuitive notion of ``sparse events'': models with bounded variability have, over every interval of fixed length, only a limited number $v$ of steps that are not stuttering (i.e., redundant repetitions).
The main result of the paper (in Section~\ref{sec:relax-dist-form}) shows how to transform efficiently any \ltl{} formula $\phi$ into a qualitative formula $\phi'$ such that $\phi$ is satisfiable \emph{over models with bounded variability} iff $\phi'$ is satisfiable over models of any variability.
The size of $\phi'$ does not depend on the distances (i.e., the number of nested occurrences of $\Xltl$) in $\phi$ but only on the maximum number of non-stuttering steps $v$.
In other words, $\phi'$ drops some information encoded in $\phi$; this information is not needed to decide satisfiability over models with bounded variability.

On the technical level, the construction that eliminates metric information relies on a normal form for \ltl{} formulas and on discrete-time generalized versions of the dense-time \emph{Pnueli operators} \cite{HR04}.
The correctness proof follows the idea of adding and removing stuttering steps to re-introduce the metric information dropped in models satisfying only qualitative constraints; it is reminiscent of the notion of \emph{stretching}, also originally introduced for dense-time models \cite{HR05,BMOW08}.
Section~\ref{sec:ltl-qualitative-ltl} first demonstrates this technique by showing how to transform any metric \ltl{} formula into a qualitative one which is equi-satisfiable (for generic models) and of polynomial size.
The feasibility of such a construction is unsurprising in hindsight, given the complexity results about qualitative \ltl{} \cite{SC85} and Etessami's construction \cite{Ete00}.
However, it is the necessary basis of the techniques used to derive the main result for models with bounded variability.

Besides the theoretical interest, the results of the present paper may be practically useful to simplify the temporal-logic analysis of systems characterized by heterogeneous components evolving over wildly different time scales, such as minutes, weeks, and years.
Assuming incommensurable distances are not a concern, such heterogeneity of \emph{time granularities} \cite{FMMR10-CSUR10} can, in principle, be modeled in terms of the finest-grain time units; but this solution comes with a significant price to pay to accommodate the largest time units in terms of the smallest, resulting in huge formulas.
If, however, the dynamics of the components with faster time scales are ``sparse'' enough, there is a redundancy in the global behavior of the system that the notion of bounded variability captures.
Hence, the analysis can be carried out more efficiently by leveraging the results of the present paper.

\paragraph{Outline.}
The paper is organized as follows.
The rest of the present section recalls related work.
Section~\ref{sec:definitions} introduces notation and basic definitions. %, including bounded variability.
Section~\ref{sec:extens-norm-forms} presents normal forms for \ltl{} formulas. % with the Pnueli operators.
Section~\ref{sec:ltl-qualitative-ltl} proves the equi-satisfiability of \ltl{} and its qualitative subset.
Section~\ref{sec:relax-dist-form} shows how the metric information can be relaxed while preserving satisfiability, for models with bounded variability.
Section~\ref{sec:future-work} concludes and outlines future work.

\subsection{Related work}
The expressiveness and complexity of \ltl{} and of its qualitative subset have been thoroughly investigated in the classic framework of temporal logic \cite{GHR94,Eme90,KMP94}.
With respect to expressiveness, Lamport introduced the notion of \emph{stuttering} to characterize qualitative \ltl{} \cite{Lam83}; the characterization was completed by Peled and Wilke \cite{PW97}, perfected by Etessami and others \cite{Ete99,Ete00,Rab98,DKL09}, and generalized by Ku{\v c}era and Strej{\v c}ek \cite{KS05}.
With respect to complexity, the seminal work of Sistla and Clarke established the PSPACE-completeness of both \ltl{} and qualitative \ltl{} \cite{SC85}, and other authors have generalized or specialized the result \cite{DS02,Mar04,BSSV08}.

To our knowledge, the present paper is the first investigating satisfiability-pre\-serv\-ing relaxations of metric information in temporal logic formulas.
More generally, the problem of formalizing systems with heterogeneous time granularities using temporal logic~\cite{FMMR10-CSUR10} has been studied by only a few authors \cite{CCM+91,DM01,CFP04,BH10}; \cite{CFP04}, in particular, presents an encoding of temporal granularities in \ltl{}, but it does not discuss efficiency of the encoding.

Some of the techniques used in the the paper borrow from existing approaches in the literature.
The normal forms for \ltl{} introduced in Section~\ref{sec:extens-norm-forms} are related to a construction used in \emph{temporal testers} \cite{PZ08-testers}.
The definition of \emph{bounded variability} in Section~\ref{sec:definitions} translates to discrete time a notion introduced for dense (or continuous) time models \cite{Wil94,Fra96,FR08-FORMATS08,CPS09}.

Hirshfeld and Rabinovich studied the expressiveness and decidability of Pnueli operators over dense time \cite{HR04}; the operators themselves were first mentioned in a conjecture attributed to Pnueli \cite{AH92b,Wil94}.
Section~\ref{sec:relax-dist-form} introduces discrete-time qualitative variants of such operators.
\emph{Counting operators}~\cite{LMP10} are somehow similar to discrete-time Pnueli operators in that they both facilitate the expression of concise counting requirements; both extensions do not increase the expressive power of \ltl{}, nor its complexity under a unary encoding.
\cite{BEH95} introduce a much more expressive counting extension of \ltl{}, which is decidable only in special cases.

The proofs of Lemmas~\ref{lem:ltl-qltl-equisat} and \ref{lem:equisat} use a technique that removes and adds stuttering steps in words to match some metric requirements; the notion of \emph{stretching} --- introduced in \cite{HR05} and further used in \cite{BMOW08} --- is similar but for dense-time models.

\section{Definitions} \label{sec:definitions}
This section introduces the syntax and semantics of \ltl{} and other basic definitions.

$\naturals$ denotes the set of natural numbers $\{0, 1, 2, \ldots\}$ and $\naturals_{>0}$ denotes the positive naturals $\{ n \in \naturals \mid n > 0 \}$.
For any two natural numbers $a \leq b$, $[a..b]$ denotes the interval of naturals $a, a+1, \ldots, b$.

\subsection{LTL formulas}

\paragraph{LTL syntax.}
The following grammar defines the set of \ltl{} formulas:
\begin{equation*}
\ltl \ni \phi ::= x \mid\neg \phi \mid \phi_1 \wedge \phi_2 \mid \phi_1 \Ultl \phi_2 \mid \Xltl \phi
\end{equation*}
where $x$ ranges over a set $\Pcal = \{p, q, r, \ldots \}$ of propositional letters.

Assume the standard abbreviations for $\logictrue, \logicfalse, \vee, \Rightarrow, \Leftrightarrow$ and for the derived temporal operators:
\begin{itemize}
\item \emph{eventually}: $\Fltl \phi \triangleq \logictrue \Ultl \phi$;
\item \emph{always}: $\Gltl \phi \triangleq \neg \Fltl \neg \phi$;
\item \emph{release}: $\phi_1 \Rltl \phi_2 \triangleq \neg (\neg \phi_1 \Ultl \neg \phi_2)$;
\item \emph{distance} $\Xltl^{k}\phi = \underbrace{\Xltl \Xltl \cdots \Xltl}_{k} \phi$ for $k \geq 0$.
\end{itemize}

\paragraph{Size and height.}
Let $\phi$ be an \ltl{} formula.
$\Pcal(\phi) \subseteq \Pcal$ denotes the (finite) set of propositional letters occurring in $\phi$.
$|\phi|$ denotes the size of $\phi$. 
Three features determine the size of $\phi$: the size $\sizep{\phi}$ of its propositional structure; the size $\sizeU{\phi}$ of its \emph{until} subformulas; and the size $\sizeX{\phi}$ of its \emph{next} subformulas.
They are defined inductively as follows.
\begin{equation*}
\langle \sizep{\phi}, \sizeU{\phi}, \sizeX{\phi} \rangle = 
\begin{cases}
\langle 1, 0, 0 \rangle
     &  \phi = x  \\
\langle 1+\sizep{\phi'}, \sizeU{\phi'}, \sizeX{\phi'}  \rangle
     &  \phi = \neg \phi'  \\
\langle 1+\sizep{\phi_1}+\sizep{\phi_2}, \sizeU{\phi_1}+\sizeU{\phi_2}, \sizeX{\phi_1}+\sizeX{\phi_2}\rangle
     &  \phi = \phi_1 \wedge \phi_2  \\
\langle \sizep{\phi_1}+\sizep{\phi_2}, 1+\sizeU{\phi_1}+\sizeU{\phi_2}, \sizeX{\phi_1}+\sizeX{\phi_2}\rangle
     &  \phi = \phi_1 \Ultl{} \phi_2  \\
\langle \sizep{\phi'}, \sizeU{\phi'}, 1+ \sizeX{\phi'}\rangle
     &  \phi = \Xltl \phi'  \\
\end{cases}
\end{equation*}
Correspondingly, $|\phi|$ is $\sizep{\phi} + \sizeU{\phi} + \sizeX{\phi}$.

For a temporal operator $\mathsf{H} \in \{\mathsf{U}, \mathsf{X}\}$, the \emph{temporal height} (or \emph{nesting depth}) $\Hcal(\phi, \mathsf{H})$ of $\mathsf{H}$ in $\phi$ is the maximum number of nested occurrences of $\mathsf{H}$ in $\phi$.
For example, $\Hcal(\phi, \Xltl) = 0$ iff $\Xltl$ is not used in $\phi$.
$\dist{\phi}$ denotes instead the maximum number of \emph{consecutive} nested occurrences of the \emph{next} operator, that is the largest $n$ such that $\Xltl{}^n$ occurs in $\phi$; clearly, $\dist{\phi} \leq \Hcal(\phi, \Xltl)$.
Finally, $\sub{\phi}$ is the number of \emph{distinct subformulas} of the form $\Xltl{}^m \phi$ with $m \geq 1$.
Notice that $\sizeX{\phi}$ is bounded by $\dist{\phi} \cdot \sub{\phi}$, hence $|\phi|$ is in
%\begin{equation*}
$\Orm\left(\sizep{\phi} + \sizeU{\phi} + \dist{\phi}\cdot\sub{\phi}\right)$.
%\end{equation*}

$\form{\mathsf{U}^{h_1}, \mathsf{X}^{h_2}}$ denotes the fragment of \ltl{} whose formulas $\psi$ are such that $\Hcal(\psi, \mathsf{U}) \leq h_1$ and $\Hcal(\psi, \mathsf{X}) \leq h_2$.
Omit the superscript to mean that there is no bound on the temporal height of an operator.
Hence, $\formk{}{}$ is the same as all \ltl{}; $\formk{}{0} = \qltl$ denotes \emph{qualitative \ltl{}}, where no \emph{next} operator is used; and $\formk{0}{0} = \propos{\Pcal}$ denotes \emph{propositional formulas} without any temporal operator.

\begin{example} \label{ex:sizes}
Consider the two formulas:
\begin{align*}
\Gamma_1 &\triangleq  \Xltl(p \wedge \Xltl{}((p \Ultl q) \wedge \Xltl{}q))
&
\Gamma_2 &\triangleq  \Xltl{}p \wedge \Xltl{}^2(p \Ultl q) \wedge \Xltl{}^3q\,.
\end{align*}
$\Gamma_1$ and $\Gamma_2$ are semantically equivalent (see Example \ref{ex:semeq}) but syntactically different; in fact, some size parameters differ in the two formulas:
$\sizep{\Gamma_1} = \sizep{\Gamma_2} = 5$; $\sizeU{\Gamma_1} = \sizeU{\Gamma_2} = 1$; $\sizeX{\Gamma_1} = 3$, $\sizeX{\Gamma_2} = 6$; $|\Gamma_1| = 12$, $|\Gamma_2| = 9$; $\Hcal(\Gamma_1, \Ultl{}) = \Hcal(\Gamma_2, \Ultl{}) = 1$; $\Hcal(\Gamma_1, \Xltl{}) = \Hcal(\Gamma_2, \Xltl{}) = 3$; $\dist{\Gamma_1} = 1$, $\dist{\Gamma_2} = 3$; $\sub{\Gamma_1} = \sub{\Gamma_2} = 3$.
\endex
\end{example}

\paragraph{$\omega$-words.}
An \emph{$\omega$-word} (or simply \emph{word}) over a set $S$ of propositional letters  is a mapping $w: \naturals \rightarrow 2^S$ or, equivalently, a denumerable sequence $w(0) w(1) \cdots$ of elements $w(i) \subseteq S$.
The set of all $\omega$-words over $S$ is denoted by $\allwords{S}$.

For $T \subseteq S$, $w|_T$ is the projection of $w$ over $T$, defined as $w(0)|_T w(1)|_T \cdots$, where $w(i)|_T = w(i) \cap T$ for all $i \in \naturals$.
The projection is extended to sets of words as expected.

For $i,j \in \naturals$, $w_i$ denotes the suffix $w(i) w(i+1) \cdots$ of $w$; $w(i,j)$ denotes the subword of $w$ of length $j$ starting at $w(i)$ (with $w(i,0) = \epsilon$ for all $i$); and $w(\range{i}{j})$ denotes the subword $w(i)w(i+1)\cdots w(j)$ (with $w(i,j) = \epsilon$ for all $j < i$).

\paragraph{LTL semantics.}
The satisfaction relation $\models$ is defined as usual, for an \ltl{} formula $\phi$, interpreted over an $\omega$-word $w$ over $\Pcal$, at position $i \in \naturals$.
\\
\begin{tabular}{l c l}
  $w,i \models p$  &  iff  & $p \in w(i)$ \\
  $w,i \models \neg \phi$ &  iff  & $w,i \not\models \phi$ \\
  $w,i \models \phi_1 \wedge \phi_2$ &   iff  & $w,i \models \phi_1$ and $w,i \models \phi_2$ \\
  $w,i \models \phi_1 \Ultl \phi_2$  & iff & there exists $j \geq i$ such that $w,j \models \phi_2$ \\
                                          && and for all $i \leq k < j$ it is $w,k \models \phi_1$ \\
  $w,i \models \Xltl \phi$ &   iff  & $w,i+1 \models \phi$ \\
  $w \models \phi$ &   iff  & $w,0 \models \phi$ \\
\end{tabular}

% \paragraph{Strict until.}
% Note that a non-strict \emph{until} is assumed, as a strict \emph{until} would make the \emph{next} operator redundant.
% Indeed, let $w,i \models p \Uplus q$ be defined as there exists $j > i$ such that $w,j \models \phi_2$ and for all $i \leq k < j$ it is $w,k \models \phi_1$.
% Then, $\Xltl p$ is equivalent to $\left( p \Rightarrow p \Uplus p \right) \wedge \left( \neg p \Uplus p \;\wedge\; p \Rplus p \right)$.

\paragraph{Satisfiability and validity.}
$\allmodels{\phi}$ denotes the set $\{ w \in \allwords{\Pcal} \mid w \models \phi\}$ of all models of $\phi$.
$\phi$ is \emph{satisfiable} iff $\allmodels{\phi} \neq \emptyset$ and is \emph{valid} iff $\allmodels{\phi} = \allwords{\Pcal}$.
Two formulas $\phi_1, \phi_2$ are \emph{equivalent} iff $\allmodels{\phi_1} = \allmodels{\phi_2}$; they are \emph{equi-satisfiable} iff they are either both satisfiable or both unsatisfiable.

\begin{proposition}[\cite{SC85}]
Checking the satisfiability of an \ltl{} or qualitative \ltl{} formula is complete for PSPACE; it can be done in time exponential in the size of the formula.
\end{proposition}

\begin{example} \label{ex:semeq}
Consider again $\Gamma_1, \Gamma_2$ in Example~\ref{ex:sizes}.
If $\Scal$ denotes the set of words $w$ such that $p \in w(1)$, $q \in w(2)$ or $p \in w(2)$, and $q \in w(3)$, then $\allmodels{\Gamma_1} = \allmodels{\Gamma_2} = \Scal$.\endex
\end{example}

\subsection{Stuttering}

%\paragraph{Stuttering.}
A position $i \in \naturals$ is \emph{redundant} in a word $w$ iff $w(i+1) = w(i)$ and there exists a $j > i$ such that $w(j) \neq w(i)$; a redundant position is also called \emph{stuttering step}.
Conversely, a \emph{non-stuttering step (nss)} is any position $i$ such that $w(i+1) \neq w(i)$ or $w(i+j) = w(i)$ for all $j \in \naturals$.

A \emph{stutter-free} word is one without stuttering steps.
Two words $w_1, w_2$ are \emph{stutter-equivalent} (or equivalent under stuttering) iff they are reducible to the same stutter-free word by removing an arbitrary number of stuttering steps.

A set of words $W$ is \emph{closed under stuttering} (or stutter-invariant) iff for every word $w \in W$, for all words $w'$ such that $w$ and $w'$ are stutter-equivalent, $w' \in W$ too.

Recall the following fundamental results about stuttering and \ltl{}.
\begin{proposition} \label{prop:stuttering}
Closure under stutter equivalence is a necessary and sufficient condition for qualitative \ltl{} languages; that is:
\begin{itemize}
\item \cite{Lam83} $\phi \in \qltl$ implies that $\allmodels{\phi}$ is closed under stutter equivalence;
\item \cite{PW97} $W$ closed under stutter equivalence and expressible in \ltl{} implies there exists $\phi \in \qltl$ such that $\allmodels{\phi} = W$.
\end{itemize}
\end{proposition}

\subsection{Variability}
Let $W$ be a set of words and $v,k$ two positive integers.
A set of propositional letters $P \subseteq \Pcal$ has \emph{variability bounded by $v/k$ in $W$} iff: for every $w \in W$, the projection $w(i,k)|_P$ over $P$ of every subword $w(i,k)$ of length $k$ has at most $v$ nss.
$\vari{P,v/k}$ denotes the set of all words where $P$ has variability bounded by $v/k$.
Note that $\vari{P,v/k}$ is not closed under stuttering for any $v < k$.

\begin{example}[The elections] \label{ex:elections}
Consider elections that occur every four years, in one of two consecutive days.
The example is deliberately kept simple to be able to demonstrate it with the various constructions of the paper.
Proposition $q$ marks the first day of every quadrennial, hence it holds initially and then precisely every $d_4 = 365 \cdot 4 = 1460$ days. 
The elections $e$ occur once within every quadrennial; precisely they occur $d_2 = 40$ or $d_3 = 41$ days before the end of the quadrennial.
Assuming models with variability bounded by $5/1460$, the behavior is completely described by the following formula.
\begin{align}
&\  q \label{elections-l1} \\
\wedge &\ 
\Gltl \left( q \Rightarrow \Xltl \left( \neg q \wedge \neg q \Ultl q \right) \wedge \Xltl^{d_4}q \right) \label{elections-l2} \\
\wedge &\ 
\Gltl\left(q \Rightarrow \Xltl\neg \left(\neg e \Ultl q \right) \right) \label{elections-l3} \\
\wedge &\ 
\Gltl\left( e \Rightarrow \neg q \;\wedge\; \Xltl\left(\neg e \Ultl q \right)\right) \label{elections-l4} \\
\wedge &\ 
\Gltl\left(e \Rightarrow \Xltl^{d_2} q \vee \Xltl^{d_3} q\right) \label{elections-l5}
\end{align} %\label{eq:elections}
The proposition $q$ marks the beginning of every quadrennial: $q$ holds initially \frf{elections-l1} and then always \emph{at least} every $d_4$ steps \frf{elections-l2}.
The elections, marked by proposition $e$, must occur once before the next quadrennial starts \frf{elections-l3}.
They must also occur not at the beginning of a new quadrennial and at most once during the quadrennial \frf{elections-l4}; precisely, they occur $d_2$ or $d_3$ days before the end of the current quadrennial \frf{elections-l5}.
%(\ref{eq:elections}) forces $q$ to occur \emph{at least} every $d_4$ steps, and $e$ to occur once $d_2$ or $d_3$ steps before the next occurrence of $q$.
A variability of $5/1460$ makes such model tight, as it allows \emph{at most} $5$ nss over a windows of length $1460$: $2$ of them accounts for $q$ becoming true and then false again once, and the other $3$ nss mark a similar double transition of $e$. \endex
\end{example}

\section{Normal forms for LTL} \label{sec:extens-norm-forms}
This section presents two normal forms for \ltl{} where the nesting of temporal operators is limited; the results in the following sections will use these normal forms.

\subsection{Flat-next form}
An \ltl{} formula is in \emph{flat-next form} (FNF) when it is written as:
\begin{equation}
\kappa \ \wedge \Gltl \left( \bigwedge_{i = 1, \ldots, N} (x_i \Leftrightarrow \Xltl{} \pi_i )\right)
\label{eq:normalform}
\end{equation}
where $\kappa \in \qltl$, $x_i \in \Pcal$, $\pi_i \in \propos{\Pcal}$.
Clearly, $\frf{eq:normalform} \in \formk{}{1}$.

The nesting depth of the $\Xltl$ operators can always be reduced to one without affecting satisfiability or complexity.
\begin{lemma}\label{lem:normalform}
For any $\phi \in \ltl{}$ it is possible to build, in polynomial time, an equi-satisfiable formula $\eta$ in FNF such that $|\eta|$ and $|\Pcal(\eta)|$ are polynomial in $|\phi|$.
\end{lemma}
\begin{proof}
Initially, let $\Qcal = \Pcal(\phi)$ and $\phi' = \phi$.
Repeat the following two steps until $\phi' \in \qltl$, with step 1 having higher precedence than step 2:
  \begin{enumerate}
    \item Replace a sub-formula of $\phi'$ in the form $\Xltl \pi$, with $\pi \in \propos{\Qcal}$, by a fresh propositional letter $p_{\Xltl \pi}$, and add $p_{\Xltl \pi}$ to $\Qcal$.
    \item Replace a maximal qualitative sub-formula $\psi \in \qltl$ of $\phi'$ that is within the scope of some $\Xltl$ operator by a fresh propositional letter $p_{\psi}$, and add $p_{\psi}$ to $\Qcal$.
    \end{enumerate}
Define $\kappa$ as $\phi' \wedge \Gltl(\bigwedge_{\substack{p_\psi \in \Qcal \\ \psi \in \qltl}}  (p_\psi \Leftrightarrow \psi))$; and $\eta$ as $\kappa \wedge \Gltl(\bigwedge_{p_{\Xltl \pi} \in \Qcal}  (p_{\Xltl \pi} \Leftrightarrow \Xltl \pi))$.
$\eta$ is in FNF and equi-satisfiable to $\phi$.
Moreover, steps 1--2 are repeated at most a number of times proportional to $|\phi|$, hence $|\eta|$ and $|\Qcal(\eta)|$ are polynomial in $|\phi|$.
 %\qed
\end{proof}

\begin{example} \label{ex:elections-flatnext}
The following is the formula in the elections Example \ref{ex:elections} in flat-next form.
\begin{equation} \label{eq:flatnext-elections}
\underbrace{
  \left( q \wedge
  \Gltl \left( \begin{array}{l}
      (u \Leftrightarrow \neg e \Ultl q) \\
      \wedge\, 
      (v \Leftrightarrow \neg q \wedge \neg q \Ultl q ) \\
      \wedge\, (q \Rightarrow x_v \wedge x_{d_4})  \\
      \wedge \, (q \Rightarrow \neg x_u) \\
      \wedge\, (e \Rightarrow \neg q \,\wedge\, x_u) \\
      \wedge\, (e \Rightarrow x_{d_2} \vee x_{d_3})
  \end{array} \right)\right) }_{\kappa}
\wedge
  \Gltl \left( \begin{array}{l}
      (x_u \Leftrightarrow \Xltl u) \wedge (x_v \Leftrightarrow \Xltl v) \\
      \wedge (x_1 \Leftrightarrow \Xltl q)  \\
      \bigwedge_{2 \leq k \leq d_4} \left( x_k \Leftrightarrow \Xltl x_{k-1}\right)
  \end{array} \right)
\end{equation}
The first conjunct is the qualitative part $\kappa$, and $x_k$ encodes $\Xltl^k q$ for $k \geq 1$. \endex
\end{example}

\subsection{Separated-next form}
An \ltl{} formula is in \emph{separated-next form} (SNF) when it is written as:
\begin{equation}
\kappa \ \wedge \Gltl \left( \bigwedge_{i = 1, \ldots, M} (x_i \Leftrightarrow \Xltl^{\dbb(i)} \pi_i) \right)
\label{eq:normalform-X}
\end{equation}
where $\kappa \in \qltl$, $x_i \in \Pcal$, $\pi_i \in \propos{\Pcal}$, and $\dbb$ is a monotonically non-decreasing mapping $[1..M] \rightarrow \naturals_{> 0}$.

Given that the FNF is a special case of the SNF, it is obvious that any \ltl{} formula can be transformed into an equi-satisfiable SNF one in polynomial time.
The SNF, however, becomes interesting when it isolates subformulas with a nesting depth of $\Xltl$ as high as possible, as stated in the following.
%, as the rest of the paper demonstrates.
\begin{lemma}\label{lem:normalform-X}
For any $\phi \in \ltl{}$ it is possible to build, in polynomial time, an equi-satisfiable formula $\eta$ in SNF \frf{eq:normalform-X} such that $|\kappa|$, $\max_i |\pi_i|$, and $|\Pcal(\eta)|$ are in $\Orm(\sizep{\phi} + \sizeU{\phi} + \sub{\phi})$, $M = \sub{\phi}$, and $\dist{\eta} = \max_i \dbb(i) = \dbb(M) = \dist{\phi}$. 
\end{lemma}
\begin{proof}
The construction mirrors the proof of Lemma~\ref{lem:normalform}, with step 1 replaced by:
\begin{itemize}
\item[1'.] Replace a sub-formula of $\phi'$ in the form $\Xltl^n \pi$ \emph{for a maximal $n \geq 1$}\ldots
\end{itemize}
$\kappa$ introduces at most a proposition for each of the $\sub{\phi}$ maximal \emph{next}-subformulas of $\phi$ and does not otherwise increase the propositional or \emph{until} structure of $\phi$ up to constant factors.
A similar reasoning applies to the maximum size of the $\pi_i$'s, which is independent of $\dist{\phi}$.
Finally, notice that $\sizep{\phi}$ bounds $|\Pcal(\phi)|$, and $|\Pcal(\eta)|$ is no larger than $2\sub{\phi} + |\Pcal(\phi)|$.
\end{proof}

% Given a formula $\phi$ in SNF (\ref{eq:normalform-X}), $\sizeU{\phi} \triangleq |\kappa|$, $\sizeG{\phi} \triangleq M$, $\sizeP{\phi} \triangleq \max_{i} \pi_i$, and $\sizeX{\phi} \triangleq \max_i \dbb(i) = \dbb(M)$ denote respectively the until, globally, propositional, and next size of the components of $\phi$. Then, $|\phi|$ is in $\Orm(\sizeU{\phi} + \sizeG{\phi} (\sizeP{\phi} + \sizeX{\phi}))$.

\begin{example} \label{ex:model-exam}
The following formula $\Omega$ is the formula of Example \ref{ex:elections} in separated-next form, with $d_1 = d_2 = 1$, $d_3 = 40$, $d_4 = 41$, $d_5 = 1460$.
\begin{equation}
\Omega \triangleq
\underbrace{\left(
  q \wedge
  \Gltl \left( \begin{array}{l}
      (u \Leftrightarrow \neg e \Ultl q) \\
      \wedge\,
      (v \Leftrightarrow \neg q \wedge \neg q \Ultl q ) \\
      \wedge\, (q \Rightarrow x_2 \wedge x_5) 
      \wedge \, (q \Rightarrow \neg x_1) \\
      \wedge\, (e \Rightarrow \neg q \,\wedge\, x_1) %\\
      \wedge\, (e \Rightarrow x_3 \vee x_4)
  \end{array} \right) \!\!
\right)}_{\kappa_\Omega}
\wedge
  \Gltl \left( \begin{array}{l}
      (x_1 \Leftrightarrow \Xltl^{d_1} u) \\
      \wedge\, (x_2 \Leftrightarrow \Xltl^{d_2} v) \\
      \wedge\, (x_3 \Leftrightarrow \Xltl^{d_3} q) \\
      \wedge\, (x_4 \Leftrightarrow \Xltl^{d_4} q) \\
      \wedge\, (x_5 \Leftrightarrow \Xltl^{d_5} q)
  \end{array} \right)
\end{equation}
Notice that $\kappa_\Omega \in \qltl$ is the first conjunct, $|\Pcal(\Omega)| = 9$, $M_\Omega = 5$, $\dist{\Omega} = d_5$; the last one dominates over the other size parameters.
The following is a model of $\Omega$.
\begin{equation*}
  \begin{array}{ccccccccccc|c}
    1    &  2      &  3      &  4      &  \cdots  &   1420    &  1421 &  1422   & 1423   & \cdots  &  1460    &  1461  \\
\hline
    q    & \neg q  & \neg q  & \neg q  &  \cdots  & \neg q  & \neg q  & \neg q  & \neg q & \cdots  & \neg q   &  q \\
\neg e   & \neg e  & \neg e  & \neg e  &  \cdots  & \neg e  &   e     & \neg e  & \neg e & \cdots  & \neg e   & \neg e \\
 u   &  \neg u      &  \neg u      &  \neg u      &  \cdots  &  \neg u   &  \neg u      &  u      &  u     &  \cdots &  u       &  u \\
\neg v   &  v      &  v      &  v      &  \cdots  &  v      &  v      &  v      &  v     &  \cdots &  v       &  \neg v \\
  \neg x_1   &  \neg x_1    &  \neg x_1    &  \neg x_1    &  \cdots  &  \neg x_1    &  x_1    &  x_1    &  x_1    &\cdots  & x_1 &  \neg x_1 \\
   x_2   &  x_2    &  x_2    &  x_2    &  \cdots  &  x_2    &  x_2    &  x_2    &  x_2    &\cdots  & \neg x_2 &  x_2 \\
\neg x_3 & \neg x_3 & \neg x_3 & \neg x_3 &\cdots &\neg x_3 &     x_3 &\neg x_3  &\neg x_3 &\cdots  &\neg x_3 & \neg x_3 \\
\neg x_4 & \neg x_4 & \neg x_4 & \neg x_4 &\cdots &     x_4 & \neg x_4&\neg x_4  &\neg x_4 &\cdots  &\neg x_4 & \neg x_4 \\
   x_5   & \neg x_5 & \neg x_5 & \neg x_5 & \cdots& \neg x_5& \neg x_5& \neg x_5 & \neg x_5& \cdots & \neg x_5& x_5 
  \end{array}
\end{equation*}
In any model of $\Omega$ corresponding to a model of \fsrf{elections-l1}{elections-l5} with variability $5/1460$ there are at most $6$ nss over $[1..1460]$: $1,1419,1420,1421,1459,1460$, corresponding to a variability of $6/1460$. % and let $\vbb_\Omega = 6$ and $D_\Omega = \dist{\Omega}$. 
\endex
\end{example}

\section{Reducing LTL to qualitative LTL} \label{sec:ltl-qualitative-ltl}
This section shows how to transform any \ltl{} formula into an equi-satisfiable \qltl{} formula of polynomially correlated size.
The following theorem summarizes the result.

\begin{theorem}\label{th:ltl-qltl-equisat}
Given an \ltl{} formula $\phi$, it is possible to build, in polynomial time, a qualitative \ltl{} formula $\xi \in \qltl{}$ such that $\phi$ and $\xi$ are equi-satisfiable and have polynomially correlated size.
\end{theorem}

Let $\eta$ be $\phi$ in FNF; the remainder of this section shows the construction of $\xi$ from $\eta$ and proves its correctness in Lemma~\ref{lem:ltl-qltl-equisat}.
Theorem~\ref{th:ltl-qltl-equisat} then follows from Lemma~\ref{lem:normalform}.

%The construction is also the foundation for the further results of Section~\ref{sec:relax-dist-form}.
%

\paragraph{Informal presentation.}
Informally, the construction to turn an \ltl{} formula into an equi-satisfiable qualitative one 
works as follows.
Introduce a fresh propositional letter $s$. %; $s$ always alternates between true and false values.
Constrain $s$ to change truth value with any propositional letter in $\Pcal$; in other words, any nss coincides with a nss of $s$.
Then, replace any occurrence of a subformula $\Xltl p$ with a suitable \emph{until} formula that defines the value of $p$ at the next nss of $s$.
In practice, this means that a formula such as $\Xltl p$ forces $p$ to hold in the next state (with a new state of $s$) only if this is necessary, i.e., if this requires a nss.
This changes the quantitative $\Xltl p$ formula into a qualitative formula where the precise metric information is relaxed.
%This allows a sort of ``compression'' of redundant stuttering steps when no state change occurs.
%Informally, this construction is a loose discrete-time analogue of ``stretching'' \cite{HR05,BMOW08}.

\paragraph{Formal presentation.}
Formally, for an \ltl{} formula in FNF $\eta \in \formk{}{1}$ over $\Pcal = \Pcal(\eta)$, we build another formula $\xi \in \qltl$ that is equi-satisfiable to $\eta$.
To this end, let $s \not\in \Pcal$ be a fresh propositional letter.
For every propositional formula $\pi \in \propos{\Pcal}$, and every $\phi \in \ltl{}$ define:
\begin{align*}
\statech{\pi} &\quad\triangleq\quad
%\left(
                          \pi \wedge \Fltl \neg \pi  \;\Rightarrow
                                     \left( \begin{array}{c}
                                     s \Ultl (\neg \pi \wedge \neg s) \\
                                              \vee \\
                                     \neg s \Ultl (\neg \pi \wedge s) \\
                                              \vee \\
                                     \bigvee_{q \in \Pcal \setminus \{\pi\}} (q \wedge \pi) \Ultl (\neg q \wedge \pi) \\
                                              \vee  \\
                                     \bigvee_{q \in \Pcal \setminus \{\pi\}} (\neg q \wedge \pi) \Ultl (q \wedge \pi) \\
                                     \end{array} \right)
%\right) 
\\
\statechm{\Pcal} &\quad\triangleq\quad
%\left(
  \left(\bigwedge_{p\in\Pcal} \Gltl p \vee \Gltl \neg p \right)
  \Rightarrow \Gltl s
%\right)
\\
\relaxOne{\phi} &\quad\triangleq\quad s \Ultl \phi \;\vee\; \neg s \Ultl \phi \\
\relaxTwo{\phi} &\quad\triangleq\quad (\phi \wedge s \:\Rightarrow\: \neg s \Rltl \phi) \;\wedge\; (\phi \wedge \neg s \:\Rightarrow\: s \Rltl \phi) \\
\relaxX{}{\phi} &\quad\triangleq\quad \relaxOne{\phi} \wedge \relaxTwo{\phi}
\end{align*}
$\statech{\pi}$ links any transition of the truth value of $\pi$ to occur simultaneously with a transition of $s$.
$\statechm{\Pcal}$ deals with the special case where no proposition ever changes truth value.
$\relaxX{}{\pi}$ is instead essentially a qualitative relaxations of the \emph{next} operator: $w,i \models \relaxX{}{p}$ holds iff the next nss of $s$ is $j \geq i$ and $w,j+1 \models p$ holds.
In particular, if $s$ never changes truth value from position $i$ on, $w,i \models \relaxX{}{\phi}$ iff $w, i \models \Fltl \phi \wedge \left( \phi \Rightarrow \Gltl \phi\right)$, for every $\phi$.

Finally, build a qualitative formula $\xi$ from $\eta$ as:
\begin{equation*}
\xi \quad \triangleq \quad \statechm{\Pcal} \wedge \bigwedge_{p \in \Pcal} \Gltl\left( \statech{p} \wedge \statech{\neg p}\right)
                        \;\wedge\; \kappa \;\wedge\;
                        \bigwedge_{i = 1, \ldots, N} \Gltl \left( \begin{array}{c}
                                                         x_i \Rightarrow \relaxX{}{\pi_i} \\
                                                         \wedge \\
                                                         \neg x_i \Rightarrow \relaxX{}{\neg \pi_i}
                                                       \end{array}\right)
\end{equation*}

It should be clear that $|\xi|$ is in $\Orm(|\eta|^2)$.
Then, the following lemma justifies the correctness of the construction given.
\begin{lemma} \label{lem:ltl-qltl-equisat}
$\eta$ and $\xi$ are equi-satisfiable formulas.
\end{lemma}
\begin{proof}
Remind that $\Pcal(\eta) = \Pcal$ and $\Pcal(\xi) =\Pcal \cup \{s\}$.
The proof is in two parts. %; for brevity only the more interesting second part is shown (see \cite{FS-RelaxingMLTL-TR10} for the complete proof).

\paragraph{$\mathbf{SAT}(\eta) \Rightarrow \mathbf{SAT}(\xi)$.}
In the first part show that $\xi$ is satisfiable if $\eta$ is satisfiable.
Hence, assume $w \models \eta$ for some $w \in \allwords{\Pcal(\eta)}$.
Build an $x \in \allwords{\Pcal(\xi)}$ such that $x \models \xi$ as follows.
$x$ coincides with $w$ over $\Pcal(\eta)$, hence $x \models \kappa$ because $s \not\in \Pcal(\kappa)$.
In addition, $s$ is added to $x$ according to the following recursive definition: $s \in x(0)$ and, for $i > 0$, if $w(i-1) = w(i)$ then $s \in x(i) \Leftrightarrow s \in x(i-1)$, whereas if $w(i-1) \neq w(i)$ then $s \in x(i) \Leftrightarrow s \not\in x(i-1)$.
In other words, $s$ switches its truth value at nss --- except possibly for an infinite tail of constant states.

For any $p \in \Pcal(\eta)$ let us show that $x \models \statechm{\Pcal} \wedge \Gltl \left( \statech{p} \wedge \statech{\neg p} \right)$.
The proof of $x \models \statechm{\Pcal}$ is routine.
Then, let $i \in \naturals$ be such that $x,i \models p$.
If $x,i \models \Gltl p$ then trivially $x,i \models \statech{p}$, because the $\pi \wedge \Fltl \neg \pi$ is false at $i$.
Otherwise, let $j > i$ be the least integer such that $x,j \models \neg p$.
If no other proposition changes truth value over $x(\range{i}{j})$, that is if $x(\range{i}{j-1})|_{\Pcal(\eta)}$ is a sequence of stuttering steps, then $s$ switches its truth value precisely at $j$.
Hence, one of $x,i \models s \Ultl (\neg p \wedge \neg s)$ and $x,i \models \neg s \Ultl (\neg p \wedge s)$ holds.
Otherwise, there exist $q \neq p$ and $k < j$ such that either $x,i \models q$ and $x,k \models \neg q$ or $x,i \models \neg q$ and $x,k \models q$.
In the former case $(q \wedge p) \Ultl (\neg q \wedge p)$ holds at $i$, whereas in the latter case $(\neg q \wedge p) \Ultl (q \wedge p)$ holds at $i$.
Hence, if $x, i \models p$ then $x,i \models \statech{p} \wedge \statech{\neg p}$ is established.
If $x, i \models \neg p$ instead, a similar reasoning also proves that $x,i \models \statech{p} \wedge \statech{\neg p}$.
In all, $x \models \Gltl (\statech{p} \wedge \statech{\neg p})$ holds.

Finally, let us prove the last conjunct of $\xi$, for a generic $h \in [1..N]$.
Let $i \in \naturals$ such that $x,i \models x_h$: we prove that $x,i \models \relaxX{}{\pi_h}$.

Since we are assuming $\eta$, $x,i+1 \models \pi_h$ holds.
If $x,i \models s$ then clearly $x,i \models s \Ultl \pi_h$.

Now, assume that $x,i \models \pi_h$: we have to show that $\neg s \Rltl \pi_h$.
That is, for a generic $j \geq i$, either $x,j \models \pi_h$ or there exists $i \leq k < j$ such that $x,k \models \neg s$.
The goal is trivial for $j = i$, as $x,i \models \pi_h$ by assumption.
It is also trivial for $j = i+1$, as $x,i+1 \models \pi_h$ also holds.
For $j > i+1$, assume adversarially that $x,j \models \neg \pi_h$.
Notice that this implies that $x,j-1 \models \neg x_h$, hence $x_h$ changes its truth value from true to false at some $i \leq m < j-1$.
Then, $x(m) \neq x(m+1)$ is not a stuttering step, which implies that $s$ also changes its truth value at $m$.
Since $s$ is true at $i$, $s$ must be false at some $i < k \leq m+1 \leq j-1 < j$.
So $x,k \models \neg s$ which closes the current branch of the proof.

Let us now consider the case $x,i \models \neg s$ hence $x,i \models \neg s \Ultl \pi_h$.
Similarly as we did in the previous case, we can establish also that if $x,i \models \pi_h$ then $s \Rltl \pi_h$.
In all, we have shown that $x,i \models \relaxX{}{\pi_h}$.

For $i \in \naturals$ such that $x,i \models \neg x_h$, a very similar reasoning shows that $x,i \models \relaxX{}{\neg \pi_h}$.

$i$ is generic, which entails the last conjuncts of $\xi$: $x \models \Gltl (x_h \Rightarrow \relaxX{}{\pi_h})$ and $x \models \Gltl (\neg x_h \Rightarrow \relaxX{}{\neg \pi_h})$.

\paragraph{$\mathbf{SAT}(\xi) \Rightarrow \mathbf{SAT}(\eta)$.}
In the second part, show that $\eta$ is satisfiable if $\xi$ is satisfiable.
Hence, assume that $w \models \xi$ for some $w \in \allwords{\Pcal(\eta) \cup \{s\}}$.
Build an $x \in \allwords{\Pcal(\eta)}$ such that $x \models \eta$ as follows.
First, let $y$ be $w$ with all stuttering steps removed. %; we then modify $y$ inductively by adding or removing stuttering steps where appropriate.
Then, let $i \in \naturals$ be a generic position and $h \in [1..N]$; since $y \models \xi$ then in particular $y,i \models x_h \Rightarrow \relaxX{}{\pi_h}$ and $y,i \models \neg x_h \Rightarrow \relaxX{}{\neg \pi_h}$.
Let us show that $y,i \models x_h \Leftrightarrow \Xltl \pi_h$.
\begin{enumerate}
\item Assume $y,i \models x_h \wedge \relaxX{}{\pi_h}$.
  Ad absurdum, let $y, i+1 \models \neg \pi_h$.
  We now discuss two cases, whether $y,i \models \pi_h$ or $y,i \models \neg \pi_h$, and we show that in both cases we reach a contradiction, hence $y, i+1 \models \pi_h$.
  \begin{enumerate}
    \item Assume $y,i \models \pi_h$.
      Also, assume that $y,i \models s$; this is without loss of generality because $\relaxX{}{\pi_h}$ is symmetric with respect to the truth value of $s$.
      Since $\pi_h$ switches from true to false at $i$, some proposition $r \neq s$ changes its truth value at $i$.
      Hence, $\statech{r} \wedge \statech{\neg r}$ forces $s$ to also change its truth value at $i$.
      In all we have the following situation:
      \begin{equation*}
      \begin{array}{cccc}
               & x_h \\
        \cdots & \pi_h  &  \neg \pi_h  & \cdots \\
               & s      &  \neg s      &  \\
               \hline
               &  i     &  i+1        
      \end{array}
      \end{equation*}
      But then $\relaxTwo{\pi_h}$ requires in particular $\neg s \Rltl \pi_h$ to hold at $i$; this is however false because neither $y,i+1 \models \pi_h$ nor $y,i \models \neg s$.
      Hence, the contradiction.

    \item Assume $y,i \models \neg \pi_h$.
      Also, assume that $y,i \models s$; this is without loss of generality because $\relaxX{}{\pi_h}$ is symmetric with respect to the truth value of $s$.
      Note that $\relaxOne{\pi_h}$ implies that $\pi_h$ must eventually hold; let $j > i+1$ be the least instant such that $y,j \models \pi_h$.
      So, $\pi_h$ does not hold at all positions in $[i..j-1]$ and becomes true at $j$.
      From the assumption that $y$ has no stuttering steps it must be $y(i+1) \neq y(i)$.
      Hence there exists some atomic proposition $r$ that changes its truth value at $i$.
      Correspondingly, $\statech{r} \wedge \statech{\neg r}$ forces $s$ to also change its truth value at $i$.
      In all we have the following situation:
      \begin{equation*}
      \begin{array}{cccccc}
               & x_h \\
        \cdots & \neg \pi_h  &  \neg \pi_h  & \cdots &  \pi_h  & \cdots \\
               & r      &   \neg r  \\
               & s      &  \neg s      &  \\
               \hline
               &  i     &  i+1        &  &  j 
      \end{array}
      \end{equation*}
      But then $\relaxOne{\pi_h}$ cannot hold at $i$, because neither $s \Ultl \pi_h$ nor $\neg s \Ultl \pi_h$ holds at $i$.
      Hence, the contradiction.
  \end{enumerate}

  \item The proof of the other case $y,i \models \neg x_h \wedge \relaxX{}{\neg \pi_h}$ can be obtained by symmetry from the previous case.
\end{enumerate}

Since $i$ and $h$ are generic, we have established $y \models \bigwedge_{i = 1, \ldots, N} \Gltl ( x_i \Leftrightarrow \Xltl{} \pi_i )$.
In addition $w \models \kappa$ implies $y \models \kappa$ as well, because $y$ is obtained from $w$ only by removing stuttering steps and $\kappa \in \qltl$ is closed under stuttering.
Hence $x = y|_{\Pcal(\eta)}$ is a model that satisfies $\eta$.  %\qed
\end{proof}

\begin{example}
% Lemma \ref{lem:ltl-qltl-equisat} establishes that \ltl{} and \qltl are equi-satisfiable, with formulas of polynomially correlated size.
% This result is of little use to reduce the practical complexity of \ltl{} satisfiability checking.
Let $\xi(\ref{eq:flatnext-elections})$ be formula (\ref{eq:flatnext-elections}) modified according to the construction of the current section.
The proof of Lemma \ref{lem:ltl-qltl-equisat} shows that the qualitative formula $\xi(\ref{eq:flatnext-elections})$ preserves the stutter-free models of the equi-satisfiable \ltl{} formula (\ref{eq:flatnext-elections}).
On the other hand, consider a model of (\ref{eq:flatnext-elections}) with a sequence of $d_5-2$ stuttering steps $\neg q \cdots \neg q q$, such as the one in Example \ref{ex:elections-flatnext}; it corresponds to the following stutter-free model of $\xi(\ref{eq:flatnext-elections})$:
\begin{equation*}
  \begin{array}{ccccccc}
 \neg q    &    \neg q   &    \cdots    &    \neg q    &    \neg q    &    \neg q   &  q\\
 \neg x_1  &    \neg x_1  &   \cdots    &    \neg x_1   &   \neg x_1   &    x_1 & \cdots  \\
 \neg x_2  &    \neg x_2  &    \cdots    &    \neg x_2   &   x_2  &  \cdots & \cdots \\
 \neg x_3  &    \neg x_3  &    \cdots    &    x_3   &    \cdots  &  \cdots & \cdots \\
    \vdots & \vdots & \vdots & \vdots & \vdots & \vdots & \vdots \\
  \neg x_{d_5 -3}  &    x_{d_5 - 3} & \cdots & \cdots & \cdots & \cdots & \cdots 
  \end{array}
\end{equation*}
This shows that the transformation of (\ref{eq:flatnext-elections}) into $\xi(\ref{eq:flatnext-elections})$ --- and more generally of $\eta$ into $\xi$ --- does not represent the redundancy of words with bounded variability more succinctly, but merely encodes it in a different form.  \endex
\end{example}

\section{LTL with bounded variability}  \label{sec:relax-dist-form}
This section specializes the results of Section~\ref{sec:ltl-qualitative-ltl} by showing how to more succinctly encode the redundancy of stuttering steps in words with bounded variability.
The following results require \expnltl{}: an extension of \qltl{} with a qualitative variant of the Pnueli operators. % (introduced elsewhere \cite{HR04}).

Section~\ref{sec:qual-pnueli-oper} recalls ``standard'' Pnueli operators, introduces \expnltl{}, and shows that \expnltl{} has the same complexity as \ltl{}.
Then, Section~\ref{sec:relax-dist-form-1} shows how to transform any \ltl{} formula $\phi$ and a positive integer parameter $\vbb$ into an \linebreak\expnltl{} formula $\phi'$ which is satisfiable (over unconstrained words) iff $\phi$ is satisfiable over words with variability bounded by $\vbb/\dist{\phi}$ --- recall that $\dist{\phi}$ is the largest distance used in $\phi$.
The size of $\phi'$ is polynomial in $\vbb$, the \emph{number} $\sub{\phi}$ of distance sub-formulas, and the size of qualitative sub-formulas appearing $\phi$; however, the size of $\phi'$ does not depend on $\dist{\phi}$ --- the \emph{values} of distances in $\phi$. 
As a consequence, checking the satisfiability of $\phi'$ --- which can be done with standard \ltl{} algorithms --- is significantly less complex than checking the original $\phi$ whenever the distances used in $\phi$ are very large and dominate over the other size parameters.

\subsection{Pnueli operators}\label{sec:qual-pnueli-oper}
Pnueli operators have been introduced for dense-time models \cite{HR04}; this section considers their discrete-time counterparts and variations thereof.

\subsubsection{Pnueli operators}
For $k,n \in \naturals$, the \emph{Pnueli operator} $\Pn{n}{k}{}$ is a $k$-ary temporal operator with the following semantics:
\begin{equation*}
w,i \models \Pn{n}{k}{}(\phi_1, \ldots, \phi_k)
\end{equation*}
holds iff there exist $k$ positions $i+1 \leq k_1 < k_2 < \cdots < k_k \leq i+n$ such that $w,k_j \models \phi_j$ for all $1 \leq j \leq k$.
It is not difficult to show that extending \ltl{} with Pnueli operators does not affect its expressiveness or complexity (this is not the case over dense time \cite{HR04}), under the assumption of a unary enconding of the integer constants. 

\begin{example} \label{ex:qPn}
Consider the following word $w$ (nss are in bold and underlined).
\begin{equation*}
  \begin{array}{cccccccccccccc}
    \mathbf{\underline{1}}   &   2   &   3   &   4   &   5   &   \mathbf{\underline{6}}   &   \mathbf{\underline{7}}   &   8   &   9   &   \mathbf{\underline{10}}   &   11   &   \mathbf{\underline{12}}   &   \mathbf{\underline{13}}   &   14 \\
    s   & \ns   &  \ns  &  \ns  &  \ns  &  \ns  &  s    &  \ns  &   \ns &  \ns   &  s     &  s     &  \ns   &  s   \\
\neg v  &   v   &   v   &   v  &    v  &    v    &   \neg v   &   v   &  v    &    v   &    v   &  v     &   \neg v    & v \\
  q     &\neg q &\neg q &\neg q &\neg q &\neg q  &\neg q  &\neg q &\neg q &\neg q  &\neg q &\neg q  &\neg q  & q \\
\neg e  & \neg e& \neg e  & \neg e& \neg e& \neg e & \neg e & e&  e & e      & \neg e & \neg e & \neg e & \neg e \\
  \end{array}
\end{equation*}
Then, $w,1 \models \Pn{13}{4}{v,s,e,s}$ and $w,8 \models \Pn{3}{2}{\neg s, \Xltl s}$ hold.
On the contrary, $w,8 \not\models \Pn{3}{3}{\neg s,\neg s, \neg s}$ because in particular $w,8+3 \not\models \neg s$.  \endex
\end{example}

The results of the present paper are based on a qualitative version of the Pnueli operators: the \emph{qualitative extended Pnueli operators} $\exPn{n}{n_1, \ldots, n_k}{k}{}$ for $k,n \in \naturals$ and $n_1, \ldots, n_k \in \naturals \cup \{*\}$.
Their semantics is defined as follows: 
\begin{equation*}
w,i \models \exPn{n}{n_1, \ldots, n_k}{k}{\phi_1, \ldots, \phi_k}
\end{equation*}
holds iff there exist $k$ positions $i \leq k_1 < \cdots < k_k$ such that all the following hold, for all $1 \leq j \leq k$:
\begin{enumerate}
\item $k_j$ is a nss;
\item $w,k_j + 1 \models \phi_j$;
\item for $j > 1$, if $n_j \neq *$ then there are no more than $n_j$ nss between $k_{j-1}$ and $k_j-1$ (both included);
\item if $n_1 \neq *$ then there are no more than $n_1$ nss between $i$ and $k_1$ (both included);
\item there are no more than $n$ nss between $i$ and $k_k$ (both included).
\end{enumerate}
Intuitively, the qualitative extended Pnueli operators are qualitative counterparts to the standard Pnueli operators, which are further generalized by imposing an additional requirement on the relative distance of $k$ nss.
For example, if $n_1 = 1$, $\phi_1$ must hold right after the first nss that follows or is at $i$, independently of the other following $k-1$ nss.
%If $\langle n_1, \ldots, n_k \rangle$ is $\langle *^k \rangle$ then the qualitative extended Pnueli modality reduces to the corresponding qualitative Pnueli modality.

\begin{example} \label{ex:qPn2}
Consider again word $w$ from Example~\ref{ex:qPn}, where nss are in bold and underlined.
For the positions $1,6,7,13$, $w,1 \models \exPn{6}{3,2,*,3}{4}{v, \neg q, e, q}$ holds.
On the contrary, $w,1 \not\models \exPn{6}{3,2,*,1}{4}{v, \neg q, e, q}$; in fact, let $k_1, \ldots, k_4$ be the positions that match the semantics of the operator.
Then, $k_4 = 13$ as $q$ only holds at $14$, so that the last component of the constraint $\langle 3,2,*,1 \rangle$ forces $k_3$ to be $12$, the nss immediately before $13$; but $w, 12+1 \not\models e$. \endex
\end{example}

\subsubsection{\ltl{} with Pnueli operators}
\expnltl{} is the extension of qualitative \ltl{} with qualitative extended Pnueli operators.
Any \expnltl{} formula has an equi-satisfiable \qltl{} formula of polynomially correlated size that can be built in polynomial time, when the integer constants used in the Pnueli operators use a unary encoding.
Precisely, to encode a formula $\exPn{n}{n_1, \ldots, n_k}{k}{\phi_1, \ldots, \phi_k}$ introduce $n^2$ letters $\{ q_i^j \mid 1 \leq i, j \leq n \}$.
Every $q_i^j$ holds iff $\exPn{j}{n_1, \ldots, n_i}{i}{\phi_1, \ldots, \phi_i}$ does; formally, a formula of size $\Orm(n \cdot \max_i |\phi_i|)$ defines $q_i^j$ as follows, where $\Xcal^j$ abbreviates $\underbrace{\Xcal \Xcal \cdots \Xcal}_j$.
\begin{equation}
  q_i^j  \Leftrightarrow
  \begin{cases}
    \logicfalse                  &   i > j \vee n_i = 0  \\
    \relaxX{}{\phi_1}                   &   1 = i = j \wedge n_i \neq 0 \\
    \underbrace{\relaxX{}{\phi_1 \vee \relaxX{}{\phi_1 \vee \cdots}}}_{j \text{ nested }\Xcal}   &   1 = i < j \leq n \wedge n_i = * \\
    \underbrace{\relaxX{}{\phi_1 \vee \relaxX{}{\phi_1 \vee \cdots}}}_{\min(j, n_1) \text{ nested }\Xcal}   &
    \begin{array}[t]{l}
      1 = i < j \leq n \\
      \wedge\; 0 < n_i \neq *
    \end{array} \\
    \left( q^{j-1}_{i-1} \wedge \relaxX{j}{\phi_i} \right)  \vee q^{j-1}_{i}   &
               \begin{array}[t]{l}
               1 < i \leq j \leq n \;\wedge \\
               (n_i = * \vee n_i \geq j)
               \end{array}  \\
    \left( \begin{array}{l}
        q^{j-1}_{i-1} \;\wedge\; \relaxX{j}{\phi_i} \ \wedge \\
        \underbrace{\relaxX{j-n_i}{\phi_{i-1} \vee \relaxX{}{\phi_{i-1} \vee \cdots}}}_{n_i \text{ nested } \Xcal} 
        \end{array} \right)
      \vee q^{j-1}_{i}   &   \begin{array}{l}
                            1 < i \leq j \leq n \\
                            \wedge \;0 < n_i < j
                            \end{array}
  \end{cases} \label{eq:qpndef}
\end{equation}
Informally, the recursive definition \frf{eq:qpndef} works as follows (from top to bottom):
\begin{itemize}
\item If $n_i = 0$ or $i > j$, $q_i^j$ is unsatisfiable.
\item If there is a positive numeric constraint $n_1 \neq *$ on when $\phi_1$ must occur, then $q_1^1$ is equivalent to $\relaxX{}{\phi_1}$: $\phi_1$ holds exactly after the next nss.
\item If $n_1 = *$ and any $j > 1$, $q_1^j$ requires that $\phi_1$ holds for some of the following $j$ nss.
\item If $n_1 \neq *$ but $n_1 \neq 0$, and $j > 1$, $q_1^j$ also requires that $\phi_1$ holds correspondingly to some of the following $n_1$ nss; if, however, $j < n_1$, then the requirement on the following next $j$ nss prevails over the constraint $n_1$.
\item If $i > 1$ and $n_i = *$ or $n_i \geq j$, the constraint $n_i$ is subsumed by the constraint $j$; then, $q_i^j$ holds iff: (a) the more constraining $q_i^{j-1}$ holds (that is, $\phi_1, \ldots, \phi_i$ hold over the following $j-1$ nss), or (b) $\phi_i$ holds exactly after the $j$-th nss, and $q_{i-1}^{j-1}$ holds as well (that is, $\phi_1, \ldots, \phi_{i-1}$ hold over the following $j-1$ nss).
\item If $i > 1$ and $n_i < j$ then $q_i^j$ is reducible to two cases; either the more constraining $q_i^{j-1}$ holds, or all of the following hold: (a) $q_{i-1}^{j-1}$ holds (which takes care of the first $i-1$ arguments $\phi_1, \ldots, \phi_{i-1}$), (b) $\phi_i$ holds exactly after the $j$-th nss, and (c) there are no more than $n_i$ nss between an occurrence of $\phi_{i-1}$ and the occurrence of $\phi_i$.
\end{itemize}

\begin{example}
Continuing Example \ref{ex:qPn2}, $w,1 \models \exPn{6}{3,2,*,3}{4}{v, \neg q, e, q}$ is rewritten as $w,1 \models q^6_4$. This in turns reduces to checking the following sequence of formulas at position $1$: $q^5_3 \wedge \relaxX{6}{q} \wedge \relaxX{3}{\relaxX{3}{q}}$; $q^4_3$; $q^3_3$; $q_2^2 \wedge \relaxX{3}{e}$; $q_1^1 \wedge \relaxX{2}{\neg q}$; $\relaxX{}{v}$. \endex
\end{example}

We can transform \expnltl{} formulas into equi-satisfiable qualitative formulas using definition \frf{eq:qpndef}.
The construction is general, but the remainder will use an \expnltl{} formula $\Lambda$ over $\Qcal = \Pcal \cup \{s\}$ in the form:
\begin{equation}
\Lambda \triangleq \statechm{\Pcal} \wedge
\bigwedge_{p \in \Pcal}\Gltl \left(\!\!\! \begin{array}{c}
    \statech{p} \wedge\\
    \statech{\neg p}
    \end{array} \!\!\!\right) \wedge \kappa \ \wedge \!\!\bigwedge_{i = 1, \ldots, M}
                \Gltl \left(
                  \xi_i \Rightarrow \exPn{\jbb(i)}{\kbb(i)}{\ibb(i)}{\psi_1^i, \ldots, \psi_{\ibb(i)}^i}
                \right)
\label{eq:expn-normalform}
\end{equation}
where $\jbb,\ibb$ are two mappings $[1..M] \rightarrow \naturals_{> 0}$; $\kbb$ is a mapping $[1..M] \rightarrow (\naturals \cup \{*\})^{\ibb(i)}$; $\kappa, \psi_i \in \qltl$ for all $1\leq i \leq M$; $\xi_i \in \propos{\Pcal}$; and $s$ does not occur in $\kappa$ or in any $\psi_i^j$.

\begin{lemma} \label{lem:it3}
$\allmodels{\Lambda}$ is closed under stuttering.
\end{lemma}
\begin{proof}
A consequence of the semantics of the extended qualitative Pnueli operators --- which introduce no metric constraint --- and of the particular form of $\Lambda$ --- where only qualitative subformulas appear.
\end{proof}

\begin{lemma} \label{lem:expn-reduction}
It is possible to build, in polynomial time, a formula $\Lambda'$ such that:
\begin{enumerate}
\item \label{it1} $\Lambda' \in \qltl$;
\item \label{it2} $|\Lambda'|$ is polynomially bounded by $|\Lambda|$; 
%\item \label{it3} $\allmodels{\Lambda}$ is closed under stuttering;
\item \label{it4} $\Lambda'$ and $\Lambda$ are equi-satisfiable.
%\item \label{it5} $\allmodels{\Lambda}$ equals the projection over $\Qcal$ of $\allmodels{\Lambda'}$.
\end{enumerate}
\end{lemma}

\begin{proof}
Let $\tau\langle \underline{q},\underline{i},\underline{j},\underline{n}, \underline{n_1}, \ldots, \underline{\phi_1}, \ldots \rangle$ denote (\ref{eq:qpndef}) with $\underline{q},\underline{i},\underline{j},\underline{n}, \underline{n_1}, \ldots, \underline{\phi_1}, \ldots$ respectively replacing $q, i, j, n, n_1, \ldots, \phi_1, \ldots$.
Introduce fresh propositional letters $q[i]^j_k$ for $1 \leq i \leq M$, $1 \leq j, k \leq \jbb(i)$
and construct $\Lambda'$ from $\Lambda$ as:
\begin{multline}
\Lambda' \triangleq
 \left(\!\!\! \begin{array}{c}
 \statechm{\Pcal} \wedge \bigwedge_{p \in \Pcal}\Gltl(\statech{p} \wedge \statech{\neg p}) \\
 \wedge \quad \kappa \quad \wedge \\
 \bigwedge_{i = 1, \ldots, M}
    \Gltl ( \xi_i \Rightarrow q[i]_{\ibb(i)}^{\jbb(i)} )
  \end{array} \!\!\!\right) \\ \wedge
  \bigwedge_{ \substack{1 \leq k \leq M \\ 1 \leq i,j \leq \jbb(k)}} \Gltl \tau\left\langle q[k],i,j,\jbb(k), \kbb(k), \psi_1^k, \ldots, \psi_{\ibb(k)}^k\right\rangle
\label{eq:qpn-normalform-prime}
\end{multline}
%where $\tau(q,i,j,n,\phi_1, \ldots, \phi_h)$ denotes the definition in \frf{eq:qpndef}.

Then, facts \ref{it1}--\ref{it4} are straightforward to prove:

\begin{itemize}
\item[\ref{it1}:] is clear by construction.

\item[\ref{it2}:] The size $|\Lambda|$ of $\Lambda$ is polynomial in $|\Qcal|, I, J, K, M$, where $J, I, K$ denote $\max_i \jbb(i)$, $\max_i \ibb(i)$, and $\max_i \max \kbb(i)$, respectively.
The first subformula of $\Lambda'$:
\begin{equation}
 \statechm{\Pcal} \wedge \bigwedge_{p \in \Pcal}\Gltl(\statech{p} \wedge \statech{\neg p})
 \wedge \quad \kappa \quad \wedge
 \bigwedge_{i = 1, \ldots, M}
    \Gltl ( \xi_i \Rightarrow q[i]_{\ibb(i)}^{\jbb(i)} )
\end{equation}
clearly has size bounded by $|\Lambda|$. % + M\cdot I \cdot J$, hence also $\Orm(|\Lambda|)$.
The second subformula of $\Lambda'$:
\begin{equation}
  \bigwedge_{ \substack{1 \leq k \leq M \\ 1 \leq i,j \leq \jbb(k)}} \Gltl \tau\left\langle q[k],i,j,\jbb(k),\kbb(k), \psi_1^k, \ldots, \psi_{\ibb(k)}^k \right\rangle
\end{equation}
has size bounded by $M \cdot J^2 \cdot \Orm(J \max_{i,j} |\psi_i^j|)$, hence also polynomial in $|\Lambda|$.

% \item[\ref{it3}:] is a straightforward consequence of the semantics of the extended qualitative Pnueli operators --- which introduce no metric constraint --- and of the particular form of $\Lambda$ --- where only qualitative subformulas appear.
%\item[\ref{it4}:] follows from \ref{it5}.

\item[\ref{it4}:] follows from the definition in (\ref{eq:qpndef}) and the semantics of the $\Xcal$ operator, along the lines of Lemma \ref{lem:ltl-qltl-equisat}. \qedhere
\end{itemize}
\end{proof}

\subsection{Relaxing distance formulas} \label{sec:relax-dist-form-1}
This section proves the following. 

\begin{theorem} \label{th:relaxing-distance}
Given an \ltl{} formula $\phi$ and an integer parameter $\vbb > 0$, it is possible to build, in polynomial time, a qualitative \ltl{} formula $\phi''$ such that:
\begin{itemize}
\item $|\phi''|$ is polynomial in $\vbb, \sizep{\phi}, \sizeU{\phi}, \sub{\phi}$ but is independent of $\dist{\phi}$;
\item $\phi$ is satisfiable over words in $\vari{\Qcal,\vbb/\dist{\phi}}$ iff $\phi''$ is satisfiable over unconstrained words.
\end{itemize}
\end{theorem}
Let $\eta$ be $\phi$ in SNF; the following construction builds a $\phi' \in \expnltl{}$ from $\eta$ such that Lemmas~\ref{lem:equisat} and \ref{lem:size} hold.
Theorem~\ref{th:relaxing-distance} follows after transforming $\phi'$ into $\phi''$ by eliminating the qualitative extended Pnueli operators according to Lemma \ref{lem:expn-reduction}. %, and by properties of $\eta$ with respect to $\phi$ (Lemma~\ref{lem:normalform-X}).

\paragraph{Informal presentation.}
Let us first informally sketch the ideas behind the transformation from $\eta$ to $\phi'$, with the aid of a few examples referring to word $w$ in Example~\ref{ex:qPn} and formula $\Omega$ in Example~\ref{ex:model-exam}.

The basic idea consists of relaxing every distance formula $\Xltl^d a$ into a qualitative formula $\relaxX{d'}{a}$ with $d' \leq \vbb$, so that consecutive nss take the role of consecutive positions.
The elimination or addition of stuttering steps reconciles words in the quantitative and qualitative transformed formulas.
For example, $w,1 \models \relaxX{6}{q}$ holds because $q$ holds at position $14$; adding $41 - 14 = 27$ repetitions of position 2 transforms $w$ into a word $w'$ where the quantitative requirement $w', 1 \models \Xltl^{41} q$ holds as well. 

The transformation must also preserve the ordering among events: if $\Xltl^d a$ and $\Xltl^e b$ both hold for some $d < e$, then $\relaxX{d'}{a}$ and $\relaxX{e'}{b}$ should hold for suitable $d' < e'$.
Another constraint requires that $e' - d' \leq e - d$; otherwise, the transformed formula admits words with $e' - d' > e - d$ non-stuttering steps between consecutive occurrences of $a$ and $b$, which may not be removable to put $a$ and $b$ at an absolute distance of $e - d$. 
For example, $\relaxX{}{s} \wedge \relaxX{2}{q}$ is a suitable relaxation of $\Xltl^{30}s \wedge \Xltl^{31}q$, whereas $\relaxX{}{s} \wedge \relaxX{3}{q}$ is not: $w,10 \models \relaxX{}{s} \wedge \relaxX{3}{q}$ but the nss 13 makes it impossible to pad $w$ with stuttering steps such that $s$ and $q$ hold at positions $10+30$ and $10+31$.

Using these ideas, a formula $\eta$ in SNF \frf{eq:normalform-X} is transformed by replacing the distance formulas with qualitative ``snapshots'' using the qualitative extended Pnueli operators: the predicates $\pi_1, \pi_2, \ldots, \pi_M$ hold orderly over some of the the following $\vbb$ nss, with the additional constraint that, between any two consecutive $\pi_i, \pi_{i+1}$, no more nss than the difference of the corresponding distances occur.
For example, if $\neg x_1 \wedge x_2 \wedge \neg x_3 \wedge \neg x_4 \wedge x_5$ (corresponding to predicates $\neg u, v, \neg q, \neg q, q$) then formula $\exPn{6}{1,*,1,*}{4}{}(\neg u \wedge v, \neg q, \neg q, q)$ must hold, where $\neg u \wedge v$ occurs after the next nss and $\neg q, \neg q$ occupy consecutive nss.

This approach can be made rigorous, but introduces an exponential blow-up because it considers each of the $2^M$ subset of propositions $x_1, \ldots, x_M$.
The following construction avoids this blow-up by introducing auxiliary propositions $y_i$'s and $z_i^j$'s that mark nss and decouple them from the propositions that must hold therein.

Each $y_i$ holds precisely from the $i$-th nss until the next $1 + (i \bmod \vbb)$ nss.
Then, for each given $h$, the propositions $z_1^h, z_2^h, \ldots, z_m^h$ (where $m$ is the number of different distances used in $\eta$) hold sequentially and cyclically from when $y^h$ holds.
Each $z_k^h$ marks a position in the sequence that satisfies the qualitative extended Pnueli operator under consideration; correspondingly, for each index $k'$ corresponding to a distance with index $k$, $\pi_{k'}$ holds with $z_k^h$ iff $x_{k'}$ holds with $y_h$.
For example, if $\neg x_1 \wedge x_2 \wedge \neg x_3 \wedge \neg x_4 \wedge x_5$ holds when some $y_{k}$ holds, then the corresponding predicates $\neg u \wedge v, \neg q, \neg q, q$ hold orderly with the next occurrences of $z_1^{k}, z_2^{k}, z_3^{k}, z_4^{k}$.

The following construction formalizes these ideas.

\paragraph{Detailed construction.}
Consider a generic \ltl{} formula $\eta$ in SNF:
\begin{equation}
\eta \quad\triangleq\quad \kappa \ \wedge \bigwedge_{i = 1, \ldots, M} \Gltl ( x_i \Leftrightarrow \Xltl^{\dbb(i)} \pi_i )
\label{eq:normalform-X-2}
\end{equation}
where $\kappa \in \qltl$, $x_i \in \Rcal = \{ x_i \mid 1 \leq i \leq M \}$, $\Qcal = \Pcal \cup \Rcal$, $\pi_i \in \propos{\Qcal}$, and $\dbb$ is a monotonically non-decreasing mapping $[1..M] \rightarrow \naturals_{> 0}$.

Introduce $\vbb$ letters $\{y_i \mid 1 \leq i \leq \vbb\}$.
Formula $\nfy$ constrains $y_i$ to occur synchronously with every $i$-th nss:
\begin{equation}
\nfy \quad\triangleq\quad  y_1 \wedge
%     \left(\bigwedge_{1 \leq i \leq \vbb} \neg y_i\right)\!\!\Ultl y_1
     \bigwedge_{1 \leq i \leq \vbb}\Gltl\left(
         y_i \Rightarrow  \left(\begin{array}{l}
                          \relaxX{}{y_{1+(i \bmod \vbb)}} \\
                          \bigwedge_{j \neq i} \neg y_j \ \wedge\\
                          y_i \Ultl y_{1+(i \bmod \vbb)}
                        \end{array} \right) \right)
\end{equation}

Let $D_1, D_2, \ldots, D_m$ be the sequence of sets that partition $[1..M]$ in such a way that indices involving the same number of consecutive nested $\Xltl$'s are in the same set, and the sets appear in the sequence in increasing order of nested $\Xltl$'s; formally: $i,j \in D_k$ with $k \triangleq D(i) = D(j)$ for some $k$ iff $\dbb(i) = \dbb(j) \triangleq d_k$ (and $d_0$ is defined as $0$); and $i \in D_{k_1}$ and $j \in D_{k_2}$ with $k_1 < k_2$ implies $\dbb(i) < \dbb(j)$.

Then, introduce another $m\cdot\vbb$ letters $\{z_i^j \mid 1 \leq i \leq m, 1 \leq j \leq \vbb\}$.
At every $i$-th nss, marked by $y_i$, the sequence $z_1^j, \ldots, z_m^j$ must hold over $m$ of the following $\vbb$ nss; moreover, between each $z_j^i$ and its preceding $z_{j-1}^i$ there must be no more than $d_i - d_{i-1}$ nss, unless $d_i - d_{i-1} > \vbb - i + 1$.
After defining, for $1 \leq i \leq m$:
\begin{equation*}
\delta_i \quad\triangleq\quad
\begin{cases}
d_i - d_{i-1}  &  \text{if } d_i - d_{i-1} \leq \vbb - i + 1 \\
*             &  \text{otherwise}
\end{cases}
\end{equation*}
the qualitative extended Pnueli operators capture this behavior of the $z_i^j$'s.
\begin{equation}
\nfzpn \quad\triangleq\quad \bigwedge_{1 \leq i \leq \vbb} \Gltl \left( y_i \Rightarrow \exPn{\vbb}{\delta_1, \ldots, \delta_m}{m}{z_1^i, \ldots, z_m^i} \right)
\end{equation}

Additionally, constrain the $z_i^j$'s to hold sequentially, according to the following.
\begin{equation}
\nfzseq \triangleq
   \left( \bigwedge_{h,j \neq 1} \neg z_{h}^j \right) \!\!\Ultl z_1^1 \wedge
\bigwedge_{\substack{1 \leq j \leq \vbb \\ 1 \leq i \leq m}} \Gltl \left( \!\!
  z_i^j \Rightarrow \!\! \left(\!\! \begin{array}{l}
      \neg z_{1 + (i \bmod m)}^{1+(j \bmod \vbb)} \\
      \wedge\,\bigwedge_{h \neq i} \neg z_{h}^j
      \end{array}\!\!\right)
    \!\!\Ultl z_{1+(i \bmod m)}^j \!\!\right)
\end{equation}

Once the $z_i^j$'s and the $y_i$'s are constrained, link the $x_i$'s to the values of the $\pi_i$'s in the distance formulas. %, according to the following idea.
If some $x_i$ holds, after or at the $j$-th nss and before the $j+1$-th, then $\pi_i$ has to hold at the $k$-th position in the sequence $z_1^j, \ldots, z_m^j$, with $k = D(i)$.
\begin{equation}
\nfzpi \quad\triangleq\quad \bigwedge_{\substack{1 \leq i \leq M \\ 1 \leq j \leq \vbb}}
\Gltl \left( \begin{array}{c}
    x_i \wedge y_j \Rightarrow \neg z_{D(i)}^j\Ultl z_{D(i)}^j \wedge \pi_i \\
    \wedge \\
    \neg x_i \wedge y_j \Rightarrow \neg z_{D(i)}^j\Ultl z_{D(i)}^j \wedge \neg \pi_i \\
    \end{array} \right)
\end{equation}

Finally, combine the various $\varkappa$ formulas to transform $\eta$ into $\phi'$:
\begin{equation}
\phi' \quad\triangleq\quad
\kappa \wedge \statechm{\Qcal} \wedge \bigwedge_{p \in \Qcal}\Gltl \left(
  \begin{array}{c} \statech{p} \wedge \\ \statech{\neg p} \end{array} \right)
\wedge \nfy \wedge \nfzpn \wedge \nfzseq \wedge \nfzpi
\end{equation}

\begin{example}
In the elections example, $\vbb = 6$, $m = 4$ instantiate $\nfy$, $\nfzseq$, and $\nfzpi$. Then, $\delta_1 = \delta_3 = 1$ and $\delta_2 = \delta_4 = *$ instantiate $\nfzpn$.  \endex
% following table 
% \begin{equation*}
%   \begin{array}{ccccc}
%     i    &    \vbb_\Omega - i + 1    &   d_i   &   d_i - d_{i-1}  &  \delta_i  \\
%     \hline
%     1    &    6                     &   1     &   1              &  1        \\
%     2    &    5                     &   40    &   39             &  *        \\
%     3    &    4                     &   41    &   1              &  1        \\
%     4    &    3                     &   1460  &   1419           &  *      
%   \end{array} 
% \end{equation*}
% instantiates $\nfzpn$.  \endex
\end{example}

The correctness of the above construction and the proof of Theorem \ref{th:relaxing-distance} rely on the following two lemmas.

\begin{lemma}  \label{lem:equisat}
$\eta$ is satisfiable over words in $\vari{\Qcal, \vbb/\dist{\phi}}$ iff $\phi'$ is satisfiable over unconstrained words.
\end{lemma}
\begin{proof} %[Proof sketch]
Let $D$ be $\dist{\eta}$, which equals $\dist{\phi}$ by Lemma~\ref{lem:normalform-X}.
The proof consists of two parts.

\paragraph{$\mathbf{SAT}(\eta) \Rightarrow \mathbf{SAT}(\phi')$.}
Let $w \in \vari{\Qcal, \vbb/D}$ such that $w \models \eta$.
$w'$ adds propositions $s, y_i, z_i^j$, constrained as follows.
$s$ switches its truth value at every nss, except for possibly an infinite tail of constant values over $w$.
Exactly one of the $y_i$'s holds at every instant, and they rotate at every nss signaled by $s$.
Whenever a given $y_j$ holds, a sequence of $z_i^j$'s hold over the following $\vbb$ nss, in a sequential fashion.
Namely, let $k$ be the first step where a certain $y_j$ holds, let $h_i$ be the last non-stuttering before position $k+d_i$, and let $l_i$ be the $\delta_i$-th nss after $h_{i-1}$ (included, with $h_0 = k$); then, $z_i^j$ starts to hold at $\min(h_i,l_i,\vbb-k+1)+1$, and holds until the next $z_{i+i}^j$.
% Notice that the fresh propositions in $w'$ do not introduce new nss, hence $w'$ is also in $\vari{\Qcal \cup \{s\} \cup \{y_i \mid 1 \leq i \leq \vbb\} \cup \{z_i^j \mid 1 \leq i \leq m , 1 \leq j \leq\vbb\}, \vbb/D}$.

Once $w'$ is built, the rest of the proof follows the lines of Lemma \ref{lem:ltl-qltl-equisat}.
It is clear that $w' \models \bigwedge_{p \in \Qcal}\statechm{\Pcal} \wedge \Gltl(\statech{p} \wedge \statech{\neg p})$ and $w' \models \kappa$.
In addition, $w' \models \nfy \wedge \nfzseq$ is a consequence of the set up of the $y_j$'s and the $z_i^j$'s.
Then, let $i$ be the current generic instant and $b \subseteq [1..M]$ be a generic subset such that $\bigwedge_{i \in b}x_i \wedge \bigwedge_{i \not\in b} \neg x_i$ holds at $i$.
Hence, $w,i \models \Xltl^{\dbb(j)}\pi_j$ holds for all $j \in b$ and $w,i \models \Xltl^{\dbb(k)}\neg \pi_k$ holds for all $k \not\in b$.
The variability of $w$ --- and that of $w'$  --- is bounded by $\vbb/D$; hence, there are at most $\vbb$ nss of item $s$ over positions $i$ to $i+D$.
Let $i \leq t_1 < \cdots < t_{\vbb} \leq i+D$ be these transition instants.
There are only stuttering steps between any such two consecutive $t_i$'s, hence there exists a subset $u_1 < \cdots < u_m$ of the $t_i$'s such that $z_{i}^j$ holds at $u_i$ for all $i$'s and some unique $j$.
Now, for all $g$ such that $D(g) = i$, $\pi_g$ holds at $k+d_i$ and (at least) since the previous and until the next nss.
Because of how each $z_i^j$'s mark the stuttering positions before $k+d_i$, for every $g$ such that $D(g) = i$, $\pi_i$ must in particular hold where $z_i^j$ first holds; because $i$ is generic, $w' \models \nfzpi$ holds.
Also, if $d_i - d_{i-1} \leq \vbb - i + 1$, there are no more than $d_i - d_{i-1}$ nss between $u_{i-1}$ and $u_i$, for all $1 \leq i \leq m$ (and assuming $u_0 = d_0 = 0$); this establishes $w', i \models \exPn{\vbb}{\delta_1, \ldots, \delta_m}{m}{z_1^j, \ldots, z_m^j}$.
In all, $w' \models \phi'$ holds.

\paragraph{$\mathbf{SAT}(\phi') \Rightarrow \mathbf{SAT}(\eta)$.}
Let $w'$ be an unconstrained word in 
\begin{equation*}
\allwords{\Qcal \cup \{s\} \cup \{y_i \mid 1 \leq i \leq \vbb\} \cup \{z_i^j \mid 1 \leq i \leq m , 1 \leq j \leq\vbb\}}
\end{equation*}
such that $w' \models \phi'$. %; notice that $\phi'$ constraints $s$, the $y_i$'s, and the $s_i^j$'s in such a way that $w' \in \{ \Qcal, \vbb/D\}$ as well.
%$w$ is initially equal to $w'$; we modify it until $w \models \eta$ as well.
Initially, let $w$ be $w'$ with all stuttering steps removed; $w \models \phi'$ as well from Lemma~\ref{lem:it3}. %(\ref{it3}).
Modify $w$ as follows, until $w \models \eta$ is the case.

Let $i$ be the current generic instant and $b \subseteq [1..M]$ be a generic subset such that $\bigwedge_{i \in b}x_i \wedge \bigwedge_{i \not\in b} \neg x_i$ holds at $i$ on $w$.
The rest of the proof works inductively on $1 \leq h \leq M$; let us focus on the more interesting inductive step.

Let $i \leq t_1 < \cdots < t_{\vbb}$ be the following $\vbb$ nss of $s$ --- and hence of any proposition in $\Qcal$ as well, according to $\statechm{\Qcal} \wedge \bigwedge_{p \in \Qcal}\Gltl(\statech{p} \wedge \statech{\neg p})$.
$\nfy$ implies that a unique $y_j$ holds at $i$; correspondingly, $\nfzpn$ entails that there exists a subset of the $u_1 < \cdots < u_m$ of the sequence $t_1 < \cdots < t_{\vbb}$ such that $z_k^j$ holds at $u_k+1$ for all $1 \leq k \leq m$.
Assume $x_h$ holds at $i$ (the case of $\neg x_h$ is clearly symmetrical and is omitted), with $g = D(h)$; then, $\nfzpi$ requires that $\pi_h$ holds with $z_g^j$ at $u_g+1$.
The inductive hypothesis implies that $u_{g-1} + 1 \leq i + d_{g-1} \leq u_g$, and $\nfzpn$ and the definition of $\delta_g$ guarantee that $u_g < i + d_g$.
Correspondingly, add $\theta \triangleq i + d_g - u_g - 1$ stuttering steps at position $u_g$ in $w$.
This ``shifts'' the previous position $u_g + 1$ to the new position $i+d_g$; hence $w,i+d_g \models \pi_h$ and $i+d_g \leq d_{g+1}$ because we added only stuttering steps.
Also, $w \models \phi'$ is still the case, because Lemma~\ref{lem:it3} guarantees that the removal or addition of stuttering steps to $w$ do not affect the satisfiability of $\phi'$.
Finally, observe that we introduced no more than $m$ nss over every subword of $w$ of length $\vbb$, and $m \leq M \leq D$ because of the pigeonhole principle, hence the variability of propositions $\Qcal$ in $w$ is bounded by $\vbb/D$.

In all, induction proves that the finally modified $w$ is such that $w \models \eta$ and $w \in \vari{\Qcal, \vbb/D}$. %\qed
\end{proof}

\begin{lemma}  \label{lem:size}
$|\phi'|$ is polynomial in $\vbb,\sizep{\phi},\sizeU{\phi},\sub{\phi}$.
\end{lemma}
\begin{proof}
The size of $\phi'$ is $|\kappa| + |\Qcal|^2 + |\nfy| + |\nfzpn| + |\nfzseq| + |\nfzpi|$, up to constant multiplicative factors.
Then, $\kappa$ is unchanged from $\eta$; $|\nfy|$ is $\Orm(\vbb^3)$; $|\nfzpn|$ is $\Orm(m^2 \cdot\vbb^2)$, which is $\Orm(M^2 \cdot\vbb^2)$; $|\nfzseq|$ is $\Orm(m^3\cdot\vbb^2)$ which is $\Orm(M^3\cdot\vbb^2)$; and $|\nfzpi|$ is $\Orm(M\cdot\vbb\cdot(M+\vbb+\max_i \pi_i))$.
The statement follows by Lemma~\ref{lem:normalform-X}.
 %\qed
\end{proof}

\begin{example}
Consider the running elections example and transform $\Omega$ (Example~\ref{ex:model-exam}) into $\Omega'$ according to the above construction.
The following is a partial model for $\Omega'$, where all propositions not appearing at some position are assumed to be false there, nss are in bold and underlined, while a hat marks successors of nss.
\begin{equation*}
  \begin{array}{cccccccccccccc}
    \mathbf{\underline{\,1\,}}   &   \widehat{\,2\,}   &   3   &   4   &   5   &   \mathbf{\underline{\,6\,}}   &   \widehat{\mathbf{\underline{\,7\,}}}   &   \widehat{\,8\,}   &   9   &   \mathbf{\underline{10}}   &   \widehat{\!11}   &   \mathbf{\underline{12}}   &   \mathbf{\underline{\,\widehat{\!13}}}   &   \widehat{\!14} \\
%         & \dst  &       &       &       &       & \dst  & \dst  &       &        & \dst   &        & \dst   & \dst \\
% \hline
    s   & \ns   &  \ns  &  \ns  &  \ns  &  \ns  &  s    &  \ns  &   \ns &  \ns   &  s     &  s     &  \ns   &  s   \\
\hline
  y_1   &  y_2  &  y_2  &  y_2  &  y_2  &  y_2   & y_3   &  y_4  &   y_4 &  y_4   & y_5    &  y_5   &  y_6   &  y_1  \\
        & z_1^1 &  z_1^1& z_1^1 & z_1^1 &  z_2^1 & z_3^1  & z_3^1 &  z_3^1& z_3^1  & z_3^1  & z_3^1  &  z_3^1 &  z_4^1 \\
        &       &       &      &       &        & z_1^2  & z_2^2 &  z_2^2& z_2^2  & z_2^2  &  z_2^2 & z_2^2  &  z_3^2 \\
        &       &       &      &       &        &        & z_1^3 & z_1^3 & z_1^3  & z_1^3  & z_1^3  & z_2^3  & z_2^3 \\
        &       &       &      &       &        &       &        &      &        & z_1^4  &  z_1^4 & z_1^4  &  z_2^4 \\
 & & & & & & & & & & & & z_1^5 & z_1^5 \\ 
 & & & & & & & & & & & & & z_1^6 \\ 
\hline
  x_2\wedge{}x_5 &      &      &       &        &        & x_4   & x_3   &  x_3    & x_1\wedge{}x_3   &   x_1    &  x_1      &    x_1    & x_1 \\
\neg v  &   v   &   v   &   v  &    v  &    v    &   v   &   v   &  v    &    v   &    v   &  v     &   v    & \neg v \\
  q     &\neg q &\neg q &\neg q &\neg q &\neg q  &\neg q  &\neg q &\neg q &\neg q  &\neg q &\neg q  &\neg q  & q \\
\neg e  & \neg e& \neg e  & \neg e& \neg e& \neg e & \neg e & e&  e & e      & \neg e & \neg e & \neg e & \neg e \\
 u  &   \neg u   &   \neg u   &   \neg u  &    \neg u  &    \neg u    &   \neg u   &   \neg u   &  \neg u    &    \neg u   &    u   &  u     &   u    & u
  \end{array}
\end{equation*}
It should be clear that the model can be transformed into one satisfying $\Omega$, such as the one in Example \ref{ex:model-exam}.
For example, the metric requirement that $e$ occur once at $1460 + 1 - 40 = 1421$ can be accommodated by removing all the stuttering steps at position $8$ and by adding $1421-8 = 1413$ additional stuttering steps at position $2$.  \endex
\end{example}

% \subsection{Examples}
% \begin{example}
% Consider a set of $4$ propositions $x_1, \ldots, x_4$ such that $x_1 \wedge x_2 \wedge \neg x_3 \wedge x_4$ hold at the current instant.
% Let $\dbb(1),\dbb(2),\dbb(3),\dbb(4)$ be $2,9,10,64$ respectively, so each element in the sequence $\pi_1, \pi_2, \neg \pi_3, \pi_4$ holds at instant $2,9,10,64$ respectively, relative to the current instant.
% Assume a variability bounded by $7$ over $64$; correspondingly represent the sequence $\pi_1, \pi_2, \neg \pi_3, \pi_4$ as a sequence of $4$ nss over the next $7$.
% That is, $\qPn{7}{4}{\pi_1, \pi_2, \neg \pi_3, \pi_4}$ holds at the current instant.
% This is not enough to capture the semantics of the original sequence at times $2,9,10,64$ because it does not constrain the relative distances.
% For example, if each element in the sequence $\pi_1, \pi_2, \neg \pi_3, \pi_4$ holds at the nss \# $1,2,6,7$, respectively, $\pi_3$ may still be true at the nss \# $4$.
% No correct behavior in the ``exact'' model corresponds to this relaxed version, because the ``spurious'' truth value of $\pi_3$ cannot be eliminated by removing (or adding) stuttering steps: $\neg\pi_3$ will not hold at the next instant after $\pi_2$ does.
% To prevent these spurious behaviors, use the extended qualitative Pnueli modality: $\exPn{7}{2,*,1,*}{4}{\pi_1, \pi_2, \neg \pi_3, \pi_4}$, which forbids sequences such as the one corresponding to the nss \# $1,2,6,7$.
% \end{example}

\section{Conclusion and future work} \label{sec:future-work}
This paper investigated satisfiability-preserving transformations of \ltl{} into its qualitative subset.
For generic models, every \ltl{} formula admits an equi-satisfiable qualitative formula of polynomially correlated size.
For models with bounded variability, every \ltl{} formula admits an equi-satisfiable qualitative formula whose size does not depend on the \emph{distances} used in the original formula, where distances are defined by nested occurrences of \emph{next} operators.
Models with bounded variability can describe the behavior of systems with time granularity heterogeneity, where components evolving with wildly different time scales coexist.
Under this assumption, the result of the present paper can be leveraged to simplify the automated reasoning of temporal logic specifications.

Future work will investigate possible generalizations and consider implementations.
Concerning theoretical aspects, we will consider extensions of the results of the present paper to:
\begin{itemize}
\item \emph{subword} stuttering \cite{KS05}, where a subword is repeated multiple times, such as in the word $abc\;abc\;abc\; \cdots$;
\item B\"uchi automata and the classical linear-time model-checking problem.
\end{itemize}
%Also, adaptation of the technique to MTL will be considered.

On the practical side, we will assess the practical usefulness of the results of the present paper.
To this end, we plan to:
\begin{itemize}
\item implement a translator from \ltl{} to formulas equi-satisfiable over words with bounded variability and combine it with off-the-shelf \ltl{} satisfiability checking tools \cite{Alaska,T2P,Zot,surveyVardi};
\item formalize systems characterized by time granularity heterogeneity, in order to determine how often the assumption of ``sparse'' events is compatible with accurate models thereof.
\end{itemize}

\paragraph{Acknowledgements.} A preliminary version of this work has been presented at the 11th Italian Conference on Theoretical Computer Science (Cremona, Italy, 28--30 Sep\-tem\-ber 2009). The authors thank the conference attendees for their comments and for interesting discussions.

%%%%% COMMENT THIS FOR arXiv
%\bibliographystyle{plain}
%%%%% COMMENT THIS FOR arXiv
%\bibliography{RelaxingMetricInfoLTL-bib}

%%%%% UNCOMMENT THIS FOR arXiv

\end{document}